\documentclass[10pt,conference]{IEEEtran}

\usepackage{acronym}
\usepackage{algpseudocode}
\usepackage{amsmath, amssymb}
\usepackage{array}
\usepackage{hyperref}
\usepackage{listings}
\usepackage{pgfplots}
\usepackage{tikz}

\title{Extracting Formal Specifications to Strenghten Type Behaviour Testing}

\author{
  \IEEEauthorblockN{Dimitri Racordon}
  \IEEEauthorblockA{
    Centre Universitaire d'Informatique\\
    Univesity of Geneva\\
    Email: dimitri.racordon@unige.ch
  }
  \and
  \IEEEauthorblockN{Didier Buchs}
  \IEEEauthorblockA{
    Centre Universitaire d'Informatique\\
    Univesity of Geneva\\
    Email: dimitri.racordon@unige.ch
  }
}

\usetikzlibrary{calc, patterns}
\usetikzlibrary{decorations.pathreplacing}
\usetikzlibrary{patterns}
\usetikzlibrary{positioning}

\pgfplotsset{compat=1.12}

\tikzstyle{task}=[circle, minimum width=5mm, draw=blue!50, fill=blue!20, thick]

\tikzstyle{cpu}=[draw, minimum height=5mm, thick]
\tikzstyle{busy}=[draw, draw=blue!50, fill=blue!20]
\tikzstyle{offline}=[draw, pattern=north east lines, pattern color=red, draw=red!50]

\tikzstyle{ddnode}=[circle, minimum width=5mm, draw=blue!50, fill=blue!20, thick]
\tikzstyle{ddleaf}=[draw, minimum width=5mm, minimum height=5mm]
\tikzstyle{faded}=[pattern=north west lines, pattern color=gray!50, draw=gray!50]

\newcommand{\setof}[1]{\ensuremath{\left \{ #1 \right \}}}

\newcolumntype{C}[1]{>{\centering\let\newline\\\arraybackslash\hspace{0pt}}m{#1}}

\acrodef{adt}[ADT]{Algebraic Data Type}
\acrodef{trs}[TRS]{Term Rewriting System}

\begin{document}

\maketitle

\begin{abstract}
  Testing has become an indispensable activity of software development,
  yet writing good and relevant tests remains a quite challenging task.
  One well-known problem is that it often is impossible or unrealistic to test for every outcome,
  as the input and/or output of a program component can represent incredbly large,
  unless infinite domains.
  A common approach to tackle this issue it to only test classes of cases,
  and to assume that those classes cover all (or at least most)
  of the cases a component is susceptible to be exposed to.
  Unfortunately, those kind of assumptions can prove wrong in many situations,
  causing a yet well-tested program to fail upon a particular input.

  In this short paper, we propose to leverage formal verification,
  in particular model checking techniques,
  as a way to better identify cases for which the aforementioned assumptions do not hold,
  and ultimately strenghten the confidence one can have in a test suite.
  The idea is to extract a formal specification of the data types of a program,
  in the form of a term rewriting system,
  and to check that specification against a set of properties specified by the programmer.
  Cases for which those properties do not hold can then be identified using model checking,
  and selected as test cases.
\end{abstract}

\section{Introduction}

Although indispensable, testing is time-consuming activity that remains extremly challenging,
despite tremendous advances in techniques and supporting tools.
Because testing cannot be exhaustive in most cases,
the set of possible test cases must be reduced to a set that tests for distinct classes of inputs,
under the assumption that if a property holds for this finite number of classes,
it also does for the entire input domain.
Unfortunately, identifying these assumptions
and ultimately building appropriate test suites is a formidable challenge.
Systematic selection techniques have been proposed \cite{bernot:1991:test-theory},
but require formal specifications,
which are a luxury some non-critical industrial developments cannot afford.

On its own, formal verification \cite{hasan:2015:formal-verification} is often seen as an alternative.
Programs are modeled in some formal language,
enabling one to formally check for proofs on some given properties.
The advantage of this approach over testing is that properties are verified
against the exhaustive set of behaviours a program may expose.
Unfortunately, despite the outstanding results formal verification has yielded for the last decades,
it has seen a relatively sparse adoption in industrial software development.
State space explosion \cite{clarke:2012:sse} often appears to be the main limitation,
but the cost to understand and/or integrate formal verification into industrial processes
is yet another reason behind this unfortunate observation.
One interesting observation reveals that tools that met the most success with the industry are
those that avoid purely mathematical notations,
either in favour of visual representations (e.g. Mathlab/Simulink \cite{dabney:2004:simulink}),
or in favour of representations close to programming (e.g. Spin/Promela \cite{holzmann:2003:spin}).
It then appears that there is a need to bridge the gap between software development and formal verification,
in order to alleviate as much as possible from both worlds.

In this short paper, we propose to extract formal specifications from actual code,
so as to enable the use of formal verification techniques, namely model checking,
to identify cases for which a test may fail.
Our extraction process relies on the assumption that in most programming languages,
the programmer is provided with a collection of basic types
that she may combine with some mechanism to form more complex data types.
By providing a formal representation for those types out of the box, in the form of \acp{adt},
and translating the semantics of the actual code in the form a \ac{trs} \cite{dick:1991:term-rewriting},
we are able to automatically build a formal specification of the program,
so as to check whether or not it satisfies a set of requirements.

\section{Our approach}

As mentioned above,
most programming languages provide the programmer with a small collection of basic types
(e.g. numeric types, collections, etc.),
as well as a mechanism to combine them to create more complex data types (e.g. composition, inheritence, etc.).
Would these basic types given a formal representations,
in our case by the means of an algebraic signature and a term rewriting system,
it is possible to extract the formal specification of the types and operations from actual code.
Consider for instance the Swift\footnote{https://swift.org} code, given in Listing \ref{lst:code}.
A type \texttt{Buffer} is defined,
with two properties \texttt{capacity} and \texttt{storage}
of type \texttt{Int} and \texttt{Array<Int>} respectively.
Assuming we already have an algebraic specification for those two types,
it is easy to create one for the type \texttt{Buffer},
as a simple composition.
The signatures of the \texttt{write} and \texttt{consume} methods are almost identical to that of Swift:
\begin{align*}
  \mathtt{write}  &: \mathtt{Buffer} \times \mathtt{Int} \to \mathtt{Buffer} \cup \mathtt{BufferError}\\
  \mathtt{consume}&: \mathtt{Buffer} \to \mathtt{Buffer} \times (\mathtt{Int} \cup \setof{\mathtt{nil}})
\end{align*}
Note that a \texttt{Buffer} term appears as part of the domain and codomain of both operations.
The one in the domain is required so that the operation can access the properties of the method is manipulating,
and the one in the codomain is required so that we can represent the possible mutation of the input buffer,
which is in fact the result of transforming imperative code into functional one.
The semantics is also easy to extract in that particular example.
The \texttt{write} function first tests whether the buffer reached its maximum capacity,
raises an exception if it did or inserts the new data if it did not.
This can be represented as the following operation, in a term rewriting system:
\begin{align*}
  &\mathtt{count}(s) \not< c \implies\\
    &\mathtt{write}(\mathtt{Buffer}(c,s),d) = \mathtt{Buffer}(c,\mathtt{cons}(d,s))\\
    \vspace{5mm}
  &\mathtt{count}(s) < c \implies\\
    &\mathtt{write}(\mathtt{Buffer}(c,s),d) = \mathtt{raise}(\mathtt{BufferError.Overflow})\\
\end{align*}
The semantics of the \texttt{consume} operation is identical to that of the \texttt{popLast} method,
from the \texttt{Array<Int>} built-in type, and hence assumed to already be provided.

\begin{figure}
  \begin{lstlisting}[
    morekeywords={struct,var,mutating,func,return,guard,else,throws,throw},
    caption={Swift implementation of a buffer},
    captionpos=b,
    label={lst:code}
  ]
struct Buffer {
  var capacity: Int   = 3
  var storage : [Int] = []

  mutating func write(data: Int) throws {
    guard storage.count < capacity else {
      throw BufferError.Overflow
    }
    storage.append(data)
  }
  mutating func consume() -> Int? {
    return storage.popLast()
  }
}
  \end{lstlisting}
\end{figure}

A formal specification is not useful by itself, but can be used to formally check requirements.
In our particular example, we propose to extend Swift to express pre/post-conditions and invariants on data types,
as depicted in Listing \ref{lst:contracts}.
Equipped with both a formal specification and a set of requirements,
we can now use model checking to find cases for which our implementation does not satisfies its requirements,
which will not reveal bugs,
but may also provide us with relevant test cases
if we are able to keep a trace of the transitions that lead to a particular counter example.

\begin{figure}
  \begin{lstlisting}[
    morekeywords={protocol,when,after,throws},
    caption={Specification of semantic requirements},
    captionpos=b,
    label={lst:contracts}
  ]
protocol Buffer {
  when storage.count == capacity
    => write(data: _) throws BufferError.Overflow
  when storage.count == 0
    => consume() == .nil
  after write(data: i)
    => consume() == i
}
  \end{lstlisting}
\end{figure}

One nice advantage of our approach over traditional ones is that the requirements are expressed
with a syntax extremly close to that of the programming language.
In our particular example, we extended Swift with some new constructs,
but a less invasive alternative would be to use comments or annotations,
as it is customary in some other languages.
This means the programmer does not need to learn a new language or tool to be able to leverage formal verification.

\section{Related works}

Our work is closely related to Meyer's the \emph{design by contract approach} \cite{meyer:2002:design-by-contract}.
The programmer is provided with a way to specify \emph{contracts}
between a \emph{supplier} (i.e. a type or an interface)
and a \emph{client} (i.e. a caller) that specifies pre/post-conditions and/or invariants
on the data that is exchanged between the two.
Contracts are traditionally checked dynamically, as the code is running.
Our approach differs in the fact that we focus on statical analysis,
with the advantage that once deemed correct,
a program does not need to carry any additional information during its execution.

Our work is also related to \emph{Abstract Testing} \cite{merz:2015:abstract-testing}.
This technique proposes to replace transitional testing with \emph{abstract cases}.
A test case is no longer described as a concrete set of inputs that should yield a concrete output,
but rather as a set of input constraints that should yield an answer that satisfies other constraints.
Then, model checking can be used to prove the correctness of the system under test.
In fact, abstract testing is very close to our approach,
and only differs in the fact that it does not produces a formal specification,
using the system under test as some kind of black box.
The advantage is that while extracting the semantics of arbitrary code might be intractable in some cases,
it is easier to call existing code and observe its behaviour.
On the other hand, a complete formal semantics will yield stronger proofs,
as it may not depend on some implementation properties.

\bibliographystyle{IEEEtran}
\bibliography{IEEEabrv,references}

\end{document}